# BREATH: A Bio-Radar Embodied Agent for Tonal and Human-Aware Diffusion Music Generation

Yunzhe Wang   Xinyu Tang   Zhixun Huang   Xiaolong Yue   Yuxin Zeng

MiLM Plus, Xiaomi Inc.

{wangyunzhe, tangxinyu, huangzhixun, yuexiaolong, zengyuxin1}@xiaomi.com

## ABSTRACT

We present a multimodal system for personalized music generation that integrates physiological sensing, LLM-based reasoning, and controllable audio synthesis. A millimeter-wave radar sensor non-invasively captures heart rate and respiration rate. These physiological signals, combined with environmental state, are interpreted by a reasoning agent to infer symbolic musical descriptors, such as tempo, mood intensity, and traditional Chinese pentatonic modes, which are then expressed as structured prompts to guide a diffusion-based audio model in synthesizing expressive melodies. The system emphasizes cultural grounding through tonal embeddings and enables adaptive, embodied music interaction. To evaluate the system, we adopt a research-creation methodology combining case studies, expert feedback, and targeted control experiments. Results show that physiological variations can modulate musical features in meaningful ways, and tonal conditioning enhances alignment with intended modal characteristics. Expert users reported that the system affords intuitive, culturally resonant musical responses and highlighted its potential for therapeutic and interactive applications. This work demonstrates a novel bio-musical feedback loop linking radar-based sensing, prompt reasoning, and generative audio modeling.

## 1. INTRODUCTION AND RELATED WORKS

Recent advances in AI music generation have enabled powerful systems capable of composing symbolic scores, generating lyrics, and synthesizing audio across diverse genres. Early symbolic models like MuseGAN[1] and SongMASS[2] explored multi-track generation using GANs[3] and Transformers[4], while large-scale audio models such as AudioLM[5], MusicLM[6] and MusicGen[7] leveraged two-stage token decoding and codec-based architectures to generate coherent music from text prompts. More recently, diffusion-based[8] approaches like MeLoDy[9], MusicLDM[10], stable audio[11], Diffrhythm[12], and AudioX[13] further improved controllability and sound quality using latent denoising strategies.

Although current generation models demonstrate strong capabilities at producing music from a single prompt, they typically lack interactive, adaptive control during multi-turn creative sessions. To overcome this limitation, recent work has explored agent-based systems that integrate explicit reasoning and multi-modal control into iterative music generation pipelines[14,15]. MusicAgent[16] is a tool-oriented LLM agent that automatically decomposes user requests into subtasks and dispatches appropriate music tools (e.g. for timbre synthesis, style conversion, classification), orchestrating multi-turn refinement while freeing users from low-level tool management. ComposerX[17] employs a multi-agent architecture in which separate agents handle melody, harmony, rhythm, and other aspects, thereby guiding GPT-4 through the composition process. Evaluations show that this cooperation yields coherent, polyphonic music that better matches user intent than single-agent approaches. ByteComposer[18] simulates a human creative pipeline via an LLM agent, combining music-theoretic knowledge and iterative feedback; outputs approach the level of novice composers in both technical and expressive quality.

These systems improve over one-shot generation by supporting interactive refinement and modular control, yet they remain largely bound to textual prompts and symbolic representations, with limited support for embodied or sensor-driven music interaction.

In parallel, there has been growing interest in sensor-based music generation, particularly involving motion, gesture, and physiological signals. Gesture-responsive systems have been proposed for tempo and phrasing control[19], and brain-computer interfaces (BCIs) have been explored for musical rhythm and intention mapping[20]. However, the use of millimeter-wave (mmWave) radar remains rare, despite its potential as a contactless, high-resolution modality for capturing heart rate, respiration, and fine motion. In this work, we explore radar as a real-time physiological interface for interactive music generation.

Another key research direction is culturally informed music generation, which remains underrepresented in large-scale generative models. While most systems rely on Western tonal frameworks (e.g., equal temperament, major/minor modes), several works such as CNPM [21] and BoYaTCN[22] have instead formalized Chinese pentatonic theory for symbolic composition. Rooted in ancient Chinese musical traditions and formalized in the 20th century by scholars such as Wang Guangqi[23] ,Li Yinghai[24], Zhengya Wang[25], this theory provides a structured framework for generating traditional-style melodies that reflect distinct tonal aesthetics. We integrate this pentatonic framework directly into the agent's reasoning and the generator's conditioning, enabling structurally grounded, culturally resonant outputs.

In this paper, we present a full-stack bio-musical interaction system combining physiological radar sensing, LLM-based reasoning, and pentatonic-tuned diffusion audio generation. At the core is a planning-capable music agent that interprets heart rate and respiration rate, captured via radar, and translates them into musical parameters (tempo, mood, instrumentation, tonality). These parameters then condition a latent diffusion model trained to synthesize high-fidelity music at 44.1 kHz. We validate our system using prompt-matching control experiments on tonal conditioning, exploratory case studies of embodied physiological-to-musical mapping, expert listener feedback sessions, and limited comparison trials assessing component effects.

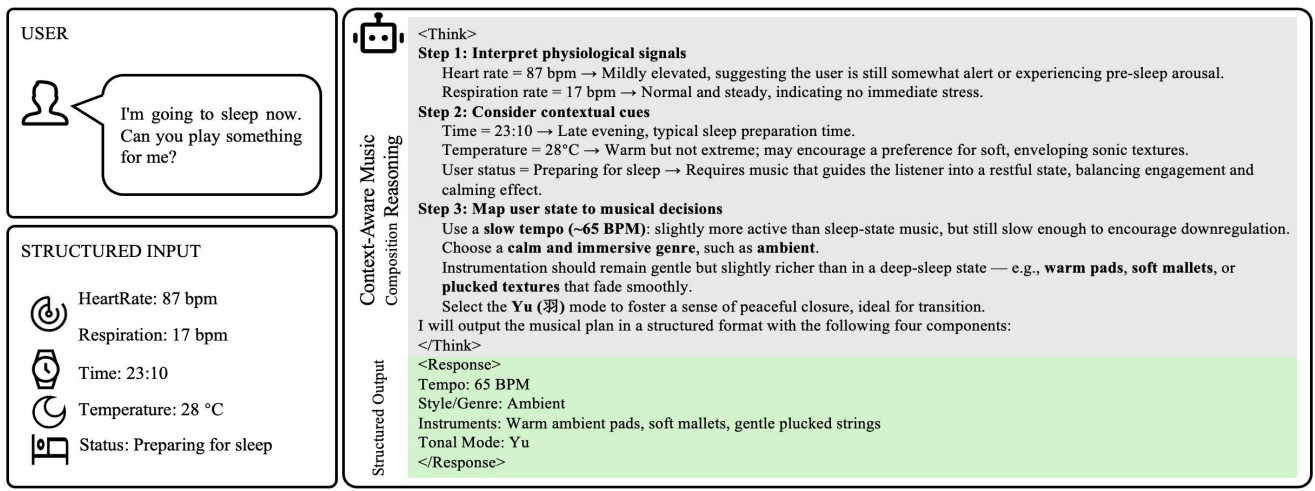

**Figure 1.** The LLM agent performs context-aware music planning by interpreting physiological and environmental inputs via step-wise symbolic reasoning. It systematically maps user state to four musical parameters—tempo, genre, instrumentation, and tonal mode—to generate personalized music supporting the transition into sleep.

Our contributions are:
- We introduce a novel multimodal agent-based system driven by millimeter-wave radar input.
- We integrate Chinese pentatonic modes as controllable tonal constraints in both planning and generation stages.
- We demonstrate an embodied LLM reasoning agent capable of mapping sensor inputs to symbolic musical structures, enabling responsive and culturally aware music generation.

## 2. SYSTEM OVERVIEW

The proposed system continuously senses a user's physiological state and generates corresponding music in real time.

At the sensing stage, our system employs a 60 GHz FMCW radar to non-invasively monitor heartbeat and respiration. Similar setups have been demonstrated in prior work: Pi-ViMo is capable of robust vital-sign separation via a multi-scatter model even under subject movement(~11-13%error)[26]; MultiVital uses mmWave MIMO radar to perform synchronized multi-point chest vibration tracking, validated against ECG/SCG references[27]. Regarding spectral processing, advanced schemes like the harmonic MUSIC algorithm achieve respiration error<3 rpm and heart-beat error<5 bpm[28], while high-precision pipelines based on DR-MUSIC and median filtering report heart-rate errors in the range of 1.7-2.6% on 77 GHz sensors[29]. These studies support the feasibility of transforming raw radar waveforms into discrete physiological states for real-time music generation via filtering and spectral analysis.

Next comes the reasoning module: a Large Language Model-based Agent. The agent takes the discretized physiological states (e.g. "heart rate high", "normal breathing") as inputs, possibly augmented by context (e.g. user preferences or prior musical context). Using a rule-informed prompt or chain-of-thought[30] reasoning, the agent determines a high-level musical plan. Concretely, it outputs a structured musical prompt specifying attributes such as tempo (BPM), style/genre, instrumentation, and tonal mode. For tonal planning, we adopt the traditional pentatonic modes (Gong, Shang, Jue, Zhi, Yu). These five modes (equivalent to scale degrees 1,2,3,5,6 in Western notation) serve as discrete conditions on melody and harmony.

Finally, the Music Generation module synthesizes audio conditioned on the agent's prompt. We employ a latent diffusion transformer (DiT) model in the style of recent audio LDMs. The prompt attributes are embedded (e.g. word embeddings for style/instrument, learned embeddings for pentatonic mode) and injected into the diffusion denoiser via cross-attention. In essence, the DiT denoises a sequence of latent audio frames to produce a high-quality waveform. The VAE[31]encoder compresses audio into a latent space, and after diffusion denoising the latent is decoded back to a 44.1kHz waveform. This path (physiological signal → agent reasoning → tonal conditioning → DiT-based audio generation) is the core workflow of our system. By design, it integrates multi-modal control: the agent can fuse physiological inputs with symbolic context or user instructions to adapt the music on-the-fly. In summary, the pipeline consists of: (1) Radar Sensing (HR/RR acquisition); (2) Signal Processing(state extraction); (3) LLM Agent (reasoning and prompt planning); (4) Diffusion Generator (DiT conditioned on musical prompt) and (5) VAE Decoder to output the final audio. This modular architecture is amenable to a block-diagram representation and supports interactive feedback, as the music adapts whenever the sensed biovariables change.

## 3. AGENT ARCHITECTURE

The LLM-based Agent serves as the system's "musical brain", mapping physiological state to symbolic music commands. Input/output: On each timestep, the agent receives a concise description of the current states (e.g. "heart rate: high", "respiration: normal", previous instrument choice, etc.) and optionally user context (e.g. "play calm music"). It then produces a textual or structured prompt specifying the desired musical attributes. In practice, we format the output as a list of parameters:

- **Tempo (BPM):** e.g. "120 BPM" or relative (fast/slow).
- **Style/Genre/Mood:** e.g. "ambient", "relaxed", or a traditional Chinese style.
- **Instrumentation:** e.g. "pads and strings".
- **Tonal mode:** one of the five pentatonic modes (Gong, Shang, Jue, Zhi, Yu)

This aligns with the notion of "music composition attributes" used in other LLM-driven systems

**Internal Reasoning.** We employ advanced prompt engineering (role-play, chain-of-thought, in-context examples) to enhance the agent's decisions. To improve consistency and interpretability, we employ chain-of-thought (CoT) prompting combined with structured role-play. For example, the agent is instructed to act as a music planner who observes heart rate, respiration, time, temperature, and user status, and generates music that supports the current state.

During inference, the agent first interprets physiological signals (e.g., heart rate = 87 bpm indicates mild arousal), then incorporates contextual cues such as time of day (e.g., 23:10 indicates pre-sleep period) and ambient temperature. Finally, it maps this user state into musical attributes such as tempo, style, instrumentation, and tonal mode using a structured and interpretable CoT format(Fig. 1).

This step-wise reasoning reflects domain-informed heuristics (e.g., "late evening + elevated heart rate → calming genre at moderate tempo") while leveraging the LLM's latent musical knowledge. Inspired by prior composition systems, we decompose the decision process into three substeps: (1) analyze physiological and environmental state; (2) decide on musical intent (e.g., transition vs. stimulation); (3) generate a structured musical response with four parameters. We find that CoT prompting enables more reliable alignment between affective input and musical output.

**Mapping to control.** The agent's output is not raw audio but control signals for the generative model. Tempo and style become conditioning embeddings input to the DiT, while the selected pentatonic mode is converted into a learned tonal embedding. For example, choosing "Gong mode" corresponds to a fixed embedding vector in the conditioning set. During inference, these embeddings guide the diffusion process: the DiT's cross-attention layers receive them as context, ensuring the generated audio follows the agent's plan. In summary, the agent architecture is:

- **Inputs**: Discrete physiological state tokens + (optional) context tokens.
- **LLM processing**: Prompted as a planner/composer, potentially using CoT or rule prompts
- **Outputs**: A structured prompt with explicit musical parameters (tempo, mode, style, instruments)

This multi-step agent loop enables the system to reason across modalities: it effectively translates body signals into symbolic music plans, which in turn condition the diffusion-based audio generator.

## 4. MUSIC GENERATION MODULE

Our music generation model follows a latent diffusion pipeline akin to the Stable Audio Tools framework. A pretrained audio VAE encoder first compresses raw audio waveforms into a lower-dimensional latent sequence. Separately, we encode textual prompts, describing instrument, style, BPM, and use-case, using a frozen pretrained T5 text encoder. The core generator is a Diffusion Transformer (DiT) operating on the latent space: at each diffusion step, a transformer block iteratively denoises the latent sequence conditioned on the prompt features. In practice, inspired by the Descript Audio Codec[32], we adopt the Stable Audio Open architecture: a fully-convolutional VAE compresses 44.1 kHz stereo into a shorter latent sequence, and the DiT ($\approx$1B parameters) uses residual and self-attention layers to predict noise conditioned on context. We train with the original latent-diffusion loss from Stable Audio, so that the model learns to reconstruct the latent audio from noise.

Crucially, we fuse both tonal and textual conditioning. Each music clip is annotated with one of the five Chinese pentatonic modes (Gong, Shang, Jue, Zhi, Yu) using a CNPM classifier. We convert this tonal label into a trainable embedding vector and tile it across all diffusion time steps. This tiled tonal embedding is concatenated to the DiT's input noise sequence, biasing generation towards the specified mode. Meanwhile, the textual prompts (instrument, style, BPM, use-case) are mapped by T5 into key-value vectors and injected via cross-attention inside each DiT block. In effect, the model attends both to the global tonal embedding and to the sequence of token embeddings from the prompt. This dual mechanism ensures that at each denoising step, the output is guided by the desired pitch mode and by semantic prompt context (e.g. "erhu in slow classical style, 60 BPM, meditation use").

During inference, we sample an initial random latent and run the DiT for many timesteps. The tonal embedding steers the melody towards the given pentatonic mode, while the cross-attended text features enforce the high-level attributes. After diffusion, the final latent sequence is passed through the pretrained VAE decoder to produce a 44.1 kHz waveform. The result is a short music clip that satisfies both sets of constraints. For example, we can ask for "slow erhu melody (style: classical) at 60 BPM for meditation" in the Gong mode, and the model will produce an audio sample reflecting all criteria.

We pre-trained the model on approximately 500K instrumental music tracks from diverse genres to learn general audio representations. For tonal conditioning, we extracted modal labels from 50K instrumental Chinese pentatonic music excerpts using a CNPM classifier, converting them into trainable embedding vectors. In the fine-tuning stage, we employed a multimodal large model to extract textual tags, such as instrument, style, tempo, and use-case, from the music excerpts, which served as prompt features during DiT training.

This combination of mode embeddings and text-based prompts is novel in audio synthesis: past text-to-audio models have only used textual or fixed conditioning. By explicitly encoding Chinese tonal class as a continuous condition alongside rich textual labels, our model achieves finer control over the generated music's melodic content and usage context. In summary, the Music Generation Module is a latent DiT that ingests (1) noise+tonal embedding and (2) text embeddings via cross-attention, and outputs latents decoded to high-fidelity audio. This design enables precise, multi-faceted conditioning of audio output in our experiments.

## 5. EXPERIMENTS

We evaluate our proposed radar-agent driven music generation system through a combination of case studies, control effectiveness analysis, and expert user feedback. Since the system is designed primarily for interactive, embodied music creation, we adopt a research-creation methodology[33] instead of conventional large-scale benchmark evaluation.

### 5.1 Case Studies: Physiological-to-Musical Mapping

We designed a series of exploratory trials in which different participants were recorded using a millimeter-wave radar sensor in a resting and lightly active state (e.g., meditative breathing, walking in place). The extracted heart rate (HR) and respiration rate (RR) values were mapped to musical conditions (tempo, mood intensity) and used to guide generation via the music agent.

For instance, in a low-HR/low-RR state, the agent selected "Gong mode, slow BPM, ambient style" as prompts and generated a calm erhu melody. Conversely, a higher RR triggered faster BPM and more intense percussive style under a "Zhi mode". In each case, the tonal conditioning (Gong, Shang, Jue, Zhi, Yu) was respected in the generated melody.

These results demonstrate that subtle changes in physiological states can be reflected in the generative system, even with minimal radar input features.

### 5.2 Effect of Tonal Embedding

To evaluate the contribution of the pentatonic tonal embedding, we generated 100 samples under three experimental conditions: (1) with tonal embedding and prompt, (2) with text prompt only, and (3) with the tonal label via natural language prompt but without embedding. An external tonal classifier was used to estimate the pentatonic mode of each output. As shown in Table 1, the tonal embedding significantly improved thealignment between the intended tonal label and the generated melody.

These results confirm that tonal embedding significantly improves the alignment between the intended tonal label and the generated melody, suggesting the effectiveness of tonal embedding conditioning.

| Experiment | Accuracy |
|---|---|
| With Tonal Embedding and Prompt | 87% |
| With Prompt Only | 22% |
| With Tonal Label in Prompt | 43% |

**Table 1.** Effect of Tonal Embedding on Tonal Accuracy

### 5.3 Expert Feedback and Use Scenarios

We invited two music technology researchers, a composer, and several non-expert users to interact with the system in a live session. Experts observed that the tonal embedding made the generated melodies "distinctively Chinese," and the responsiveness to radar input "opened up possibilities for embodied performance." Both experts and users noted that the system could help create atmospheres conducive to mental well-being, suggesting potential applications in rehabilitative or therapeutic music contexts. One composer even proposed leveraging the system for music therapy, where patient respiration could shape real-time accompaniment to support emotional regulation and relaxation.

## 6. CONCLUSION AND FUTURE WORK

We presented a multimodal music generation framework that integrates non-contact physiological sensing, LLM-based reasoning via a language-model agent, and controllable diffusion-based audio synthesis. By combining millimeter-wave radar input with a prompt-planning agent grounded in Chinese pentatonic mode theory, the system enables culturally-informed, bio-adaptive music generation. Our experiments demonstrate that even minimal physiological features, heart rate and respiration rate, can meaningfully guide musical outputs through the agent's planning. Furthermore, we show that explicit tonal embeddings and prompt conditioning significantly enhance musical coherence and user-perceived alignment with intent.

In future work, we plan to extend the system's perceptual input space by incorporating additional modalities such as motion (via IMU) and affective signals (e.g., facial expression, GSR). We also aim to strengthen the agent's reasoning capabilities with feedback-driven loops, enabling real-time co-adaptation between user state and musical structure. On the cultural modeling side, we are exploring finer-grained representations of Chinese modes, including modulation, ornamentation, and instrument-specific phrasing. Ultimately, this work contributes to a growing direction in interactive music AI that foregrounds embodiment, personalization, and cultural specificity in generative systems.